\documentclass{PoS}
\usepackage{amsmath}
\usepackage{graphicx}
\usepackage{nicefrac}
\usepackage{citesort}
\usepackage{wrapfig}

\usepackage{caption,setspace}
\usepackage{lipsum,lineno}
\usepackage[none]{hyphenat}
\makeatletter 
\renewcommand\speaker[1]{\if@speaker\global\@dblspeaktrue\fi
			\global\@speakertrue
			\global\setbox\@firstaubox
			\hbox{{\let\thanks\@gobble
				\let\footnote\@gobble\small
				\rm  Bo-Ting Wang {\it et al.}, }}%
			#1\thanks{Speaker.}\
			}%
\makeatother

\makeatletter 
\@ifundefined{textcolor}{}
{%
 \definecolor{BLACK}{gray}{0}
 \definecolor{WHITE}{gray}{1}
 \definecolor{RED}{rgb}{1,0,0}
 \definecolor{GREEN}{rgb}{0,1,0}
 \definecolor{BLUE}{rgb}{0,0,1}
 \definecolor{CYAN}{cmyk}{1,0,0,0}
 \definecolor{MAGENTA}{cmyk}{0,1,0,0}
 \definecolor{YELLOW}{cmyk}{0,0,1,0}
}

\PassOptionsToPackage{english}{babel}

\usepackage{float}
\usepackage{mathrsfs}
\usepackage{amsfonts}
\usepackage{array}
\usepackage{epsfig}

\def\beq{\begin{equation}}
\def\b0{\beta_0}

\def\eeq{\end{equation}}
\def\beeq{\begin{eqnarray}}
\def\eeeq{\end{eqnarray}}

\def\to{\rightarrow}

\makeatother

\title{\textsc{PDFSense}:\thanks{%
The webpage for the 
\textsc{\bf PDFSense} tool is: \quad
\texttt{\bf https://metapdf.hepforge.org/PDFSense/}
\null \vspace{5pt}
\newline 
We acknowledge the hospitality of CERN, DESY, and Fermilab where a
portion of this work was performed.
This work was also partially supported by the U.S.\ Department of
Energy under Grant No.\ DE-SC0010129
and by the
National Natural Science Foundation of China under the Grant No.~11465018.
\null \vspace{5pt}
}
\  Mapping the sensitivity of hadronic experiments to nucleon structure
}

\ShortTitle{\textsc{PDFSense}: Visualizing the sensitivity }

\def\thanksref#1{\rlap,${}^{#1}$}
\def\inst#1{\hangafter=1\hangindent=15pt\relax ${}^{#1}$}

\author{
Bo-Ting Wang\thanksref{a} \ 
T. J. Hobbs\thanksref{a} \ 
Sean Doyle\thanksref{a} \ 
Jun Gao\thanksref{b} \ 
Tie-Jiun Hou\thanksref{c} \ 
Pavel~M.~Nadolsky\thanksref{a} \ 
Fredrick I. Olness\thanksref{a}\speaker{} \ 
\\
\inst{a} Department of Physics, Southern Methodist University,
 Dallas, TX 75275-0181, U.S.A.  \\
\inst{b} Shanghai Key Laboratory for Particle Physics and Cosmology,
 School of Physics and Astronomy, INPAC,
 Shanghai Jiao-Tong University, Shanghai 200240, China \\
\inst{c}
School of Physics Science and Technology, Xinjiang University,
 Urumqi, Xinjiang 830046 China \\
}
\abstract{%

Recent high precision experimental data from a variety of hadronic
processes
opens new opportunities for determination of
the collinear parton distribution functions (PDFs) of the proton.
In fact, the wealth of information from experiments such as the 
Large Hadron Collider  (LHC) and others,  makes it difficult to
quickly assess the impact on the PDFs, short of performing
computationally expensive global fits. 
As an alternative, we
explore new methods for quantifying the potential impact of
experimental data on the extraction of proton PDFs. Our
approach relies crucially on the  correlation between
theory-data residuals and the PDFs themselves, as well as on a
newly defined quantity --- the {\it sensitivity} --- which
represents an extension of the correlation and reflects both
PDF-driven and experimental uncertainties. 
This approach is
realized in a new, publicly available analysis package
\textsc{PDFSense}, which operates with these statistical
measures to identify particularly sensitive experiments, weigh
their relative or potential impact on PDFs, and visualize
their detailed distributions in a space of the parton momentum
fraction $x$ and factorization scale $\mu$. 
This tool offers a
new means of understanding the influence of individual
measurements in existing fits, as well as a predictive device
for directing future fits toward the highest impact data and
assumptions.
}

\FullConference{XXVI International Workshop on Deep-Inelastic Scattering and Related Subjects (DIS2018)\\
		16-20 April 2018\\
		Kobe, Japan}

\begin{document}

\def\figone{
\begin{figure*}[t]
\centering{}
\null \vspace{-0.5cm}
\includegraphics[width=0.47\textwidth]{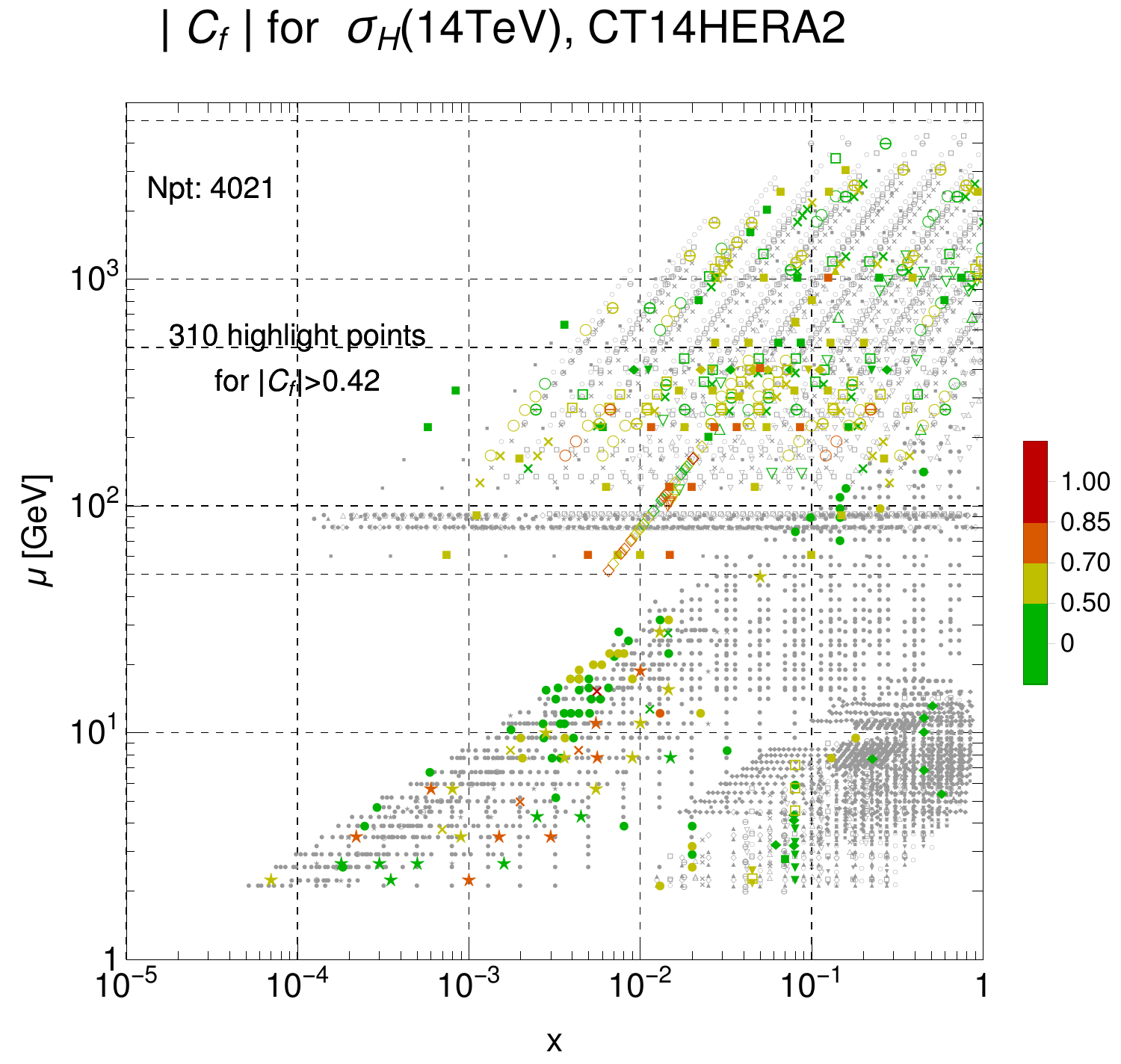}\ \ \
\includegraphics[width=0.47\textwidth]{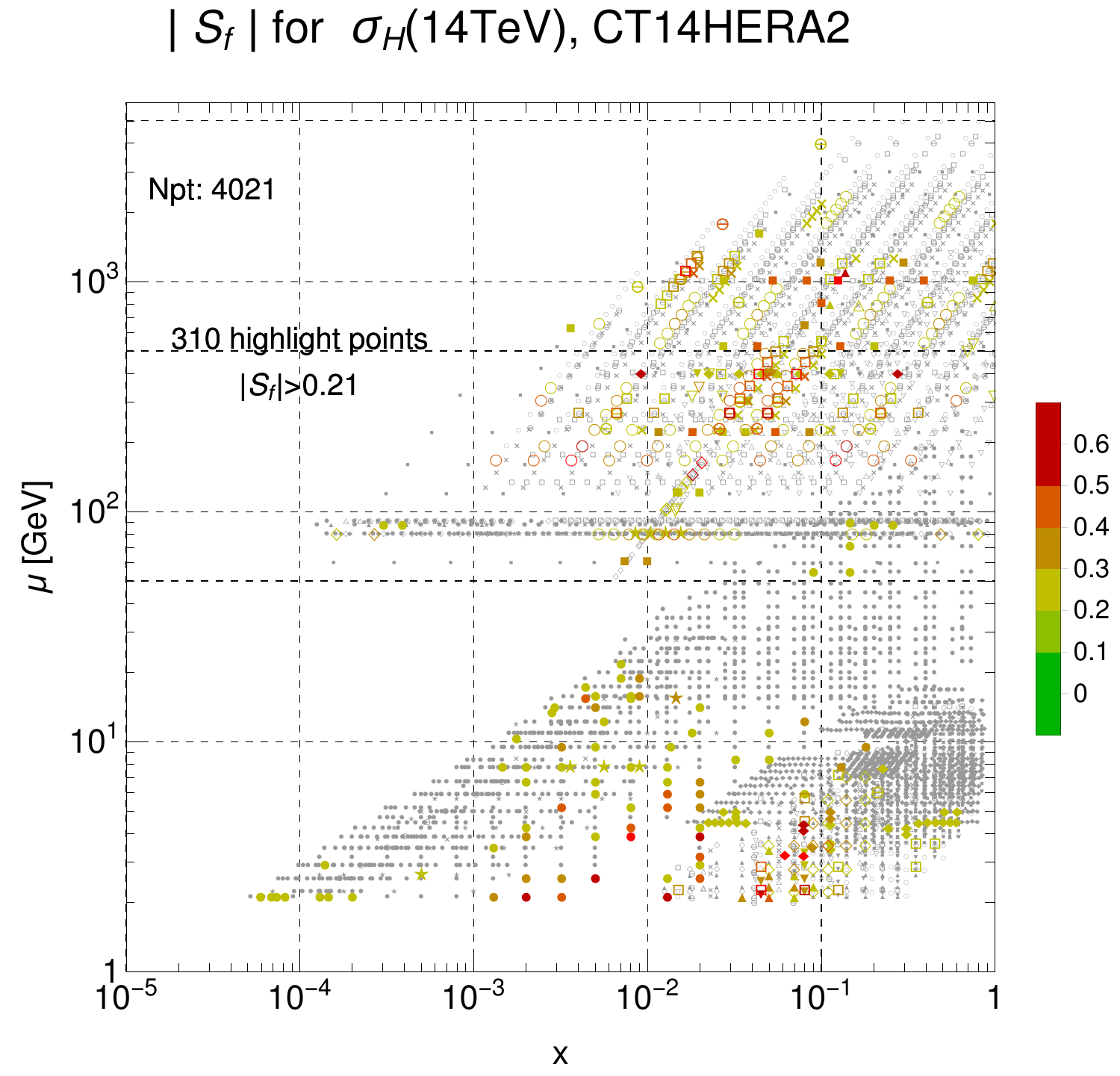}
\caption{For the full CTEQ-TEA data-set (CT14HERA2), we show the
absolute correlation $|C_{f}|$ and sensitivity $|S_{f}|$ associated
with the 14 TeV Higgs production cross section $\sigma_{H^{0}}(14\,\mathrm{TeV})$.
Points with significant magnitudes of $|C_{f}|$ and $|S_{f}|$ are
highlighted with color; the thresholds are chosen to highlight the same effective number of points (310)
in both plots. 
When  the $|C_{f}|$ plot at left is considered, only a very small
sub-population of jet production data (diagonal open circles and closed squares
with $\mu\gtrsim100$ GeV) exhibits significant correlations,
as well as some HERA DIS and $t\bar{t}$ production data points.
Conversely, the sensitivity in the right panel 
reveals a broader range of points constraining the Higgs cross section. Here,
a larger fraction of jet production points are important (especially CMS measurements),
as well as processes at smaller $\mu$, particularly
DIS data from HERA and fixed-target experiments ({\it e.g.}, BCDMS, NMC, CDHSW, and CCFR). 
Although its cumulative impact
is comparatively modest, ATLAS $t\bar{t}$
production data register significant per-point sensitivities,
as do CCFR $F^2_p$ measurements and CMS 7 TeV $A_\mu$ data;
similarly, some of the high-$p_T$ $Z$ production information
from ATLAS provide modest constraints.
\label{fig:CorrSensH14}
}
\end{figure*}
}

\def\figtwo{
\begin{figure*}
\centering{}
\null \vspace{-0.5cm}
\includegraphics[clip,width=0.415\textwidth]{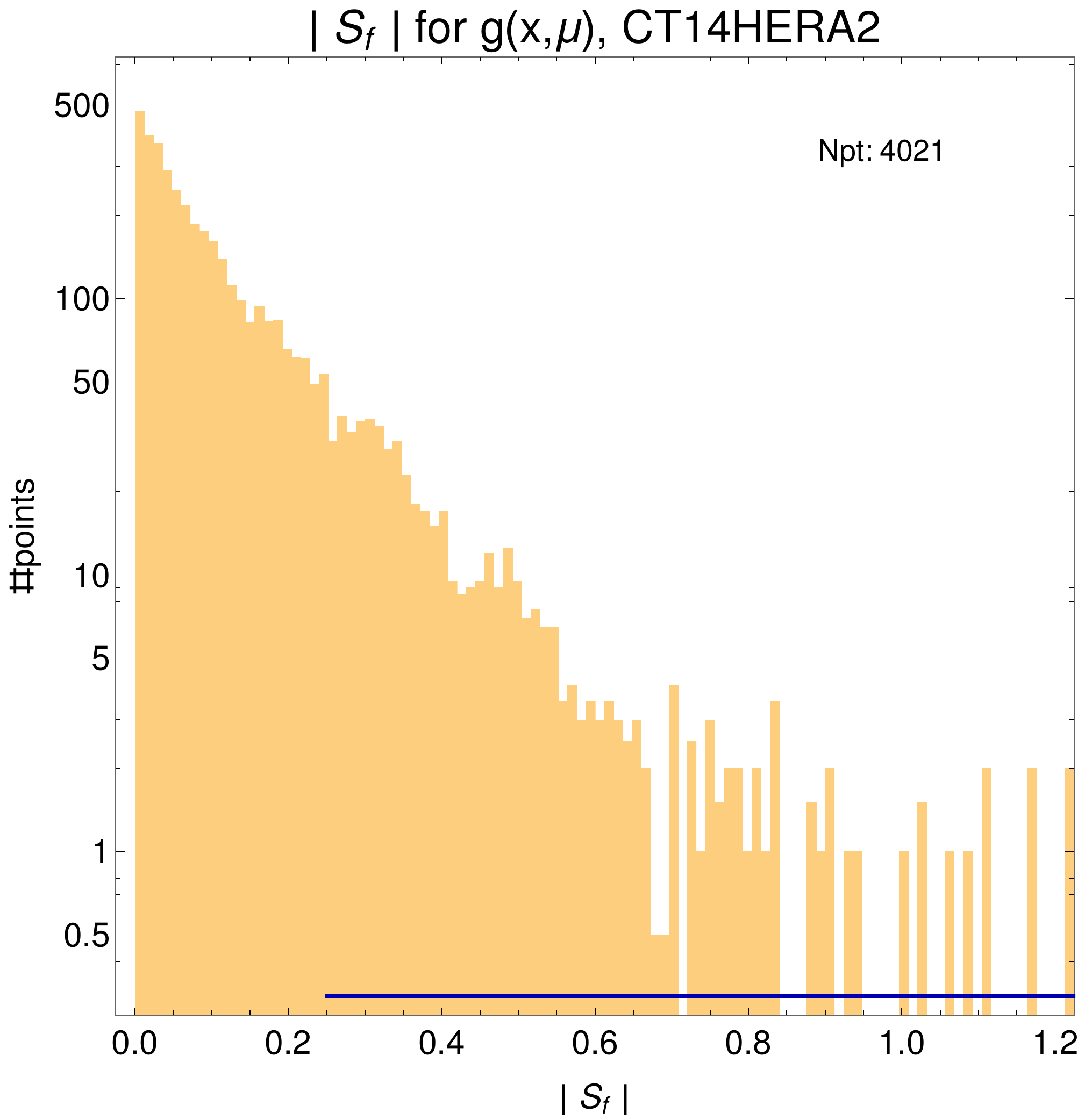}
\includegraphics[clip,width=0.47\textwidth]{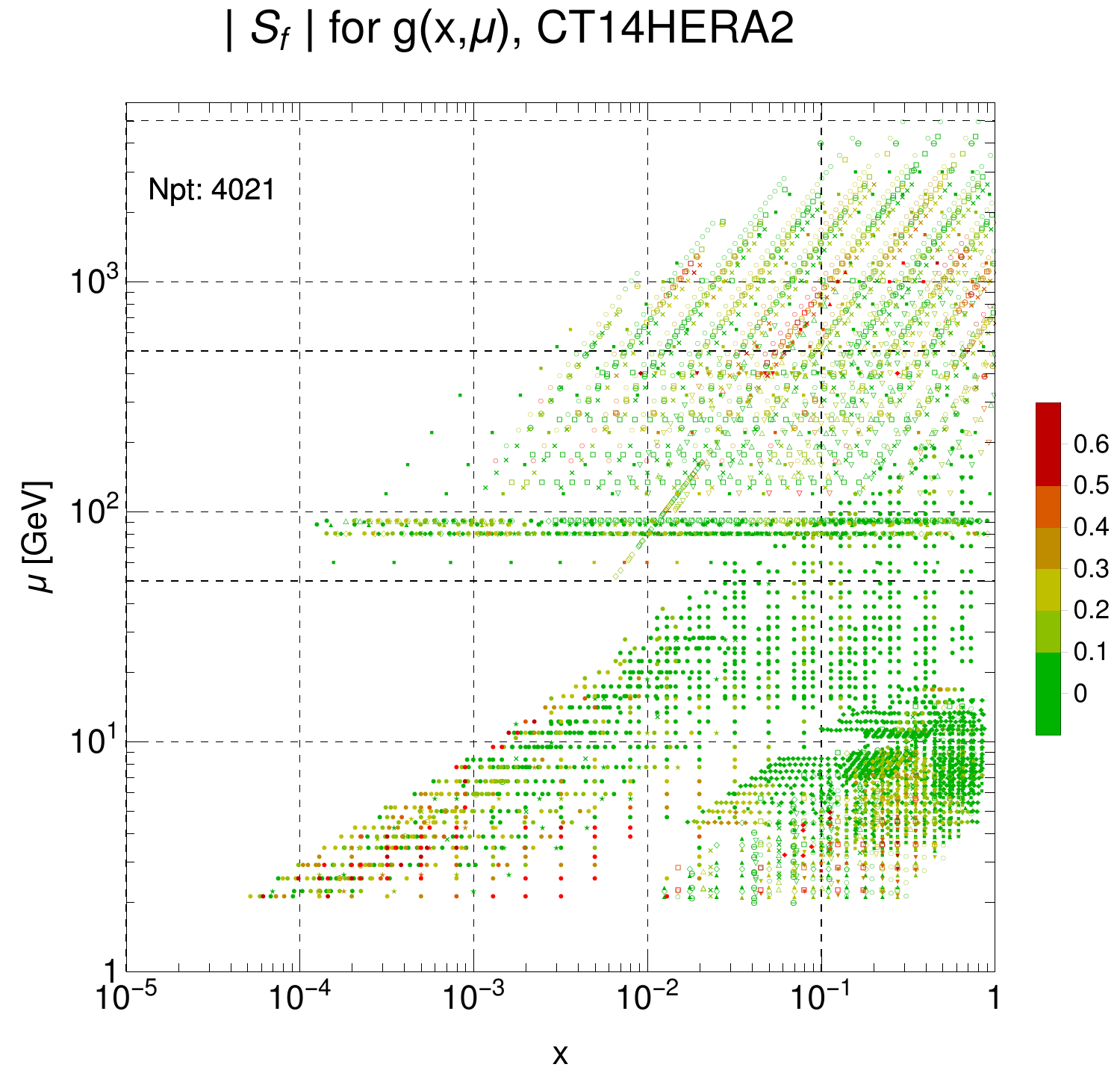}
\caption{
Two representations of the sensitivity $|S_{g}|(x_{i},\mu_{i})$ to
the gluon PDF $g(x,\mu)$ of the experimental measurements making up
the augmented CTEQ-TEA data-set; in the left panel we plot a histogram
showing the distribution of sensitivities for 4021 physical measurements.
In the right panel we show the $\{x_{i},\mu_{i}\}$ map corresponding
to these data within the full data-set.}
\label{fig:sens-main} 
\end{figure*}
}

\def\figthree{
\begin{figure*}[t]
\null \vspace{-0.5cm}
\centering{}
\includegraphics[width=0.48\textwidth]{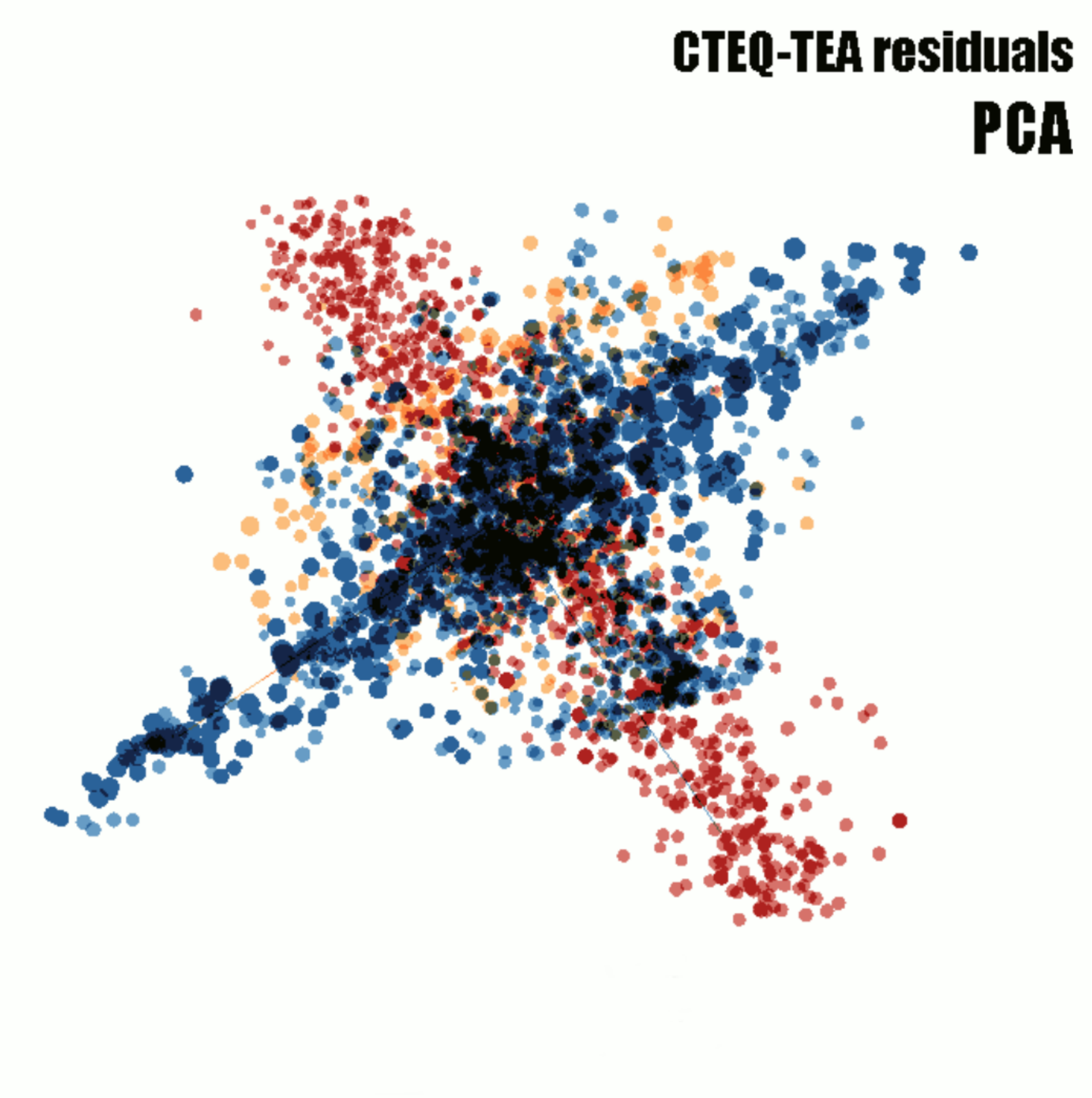}
\hfil
\includegraphics[width=0.48\textwidth]{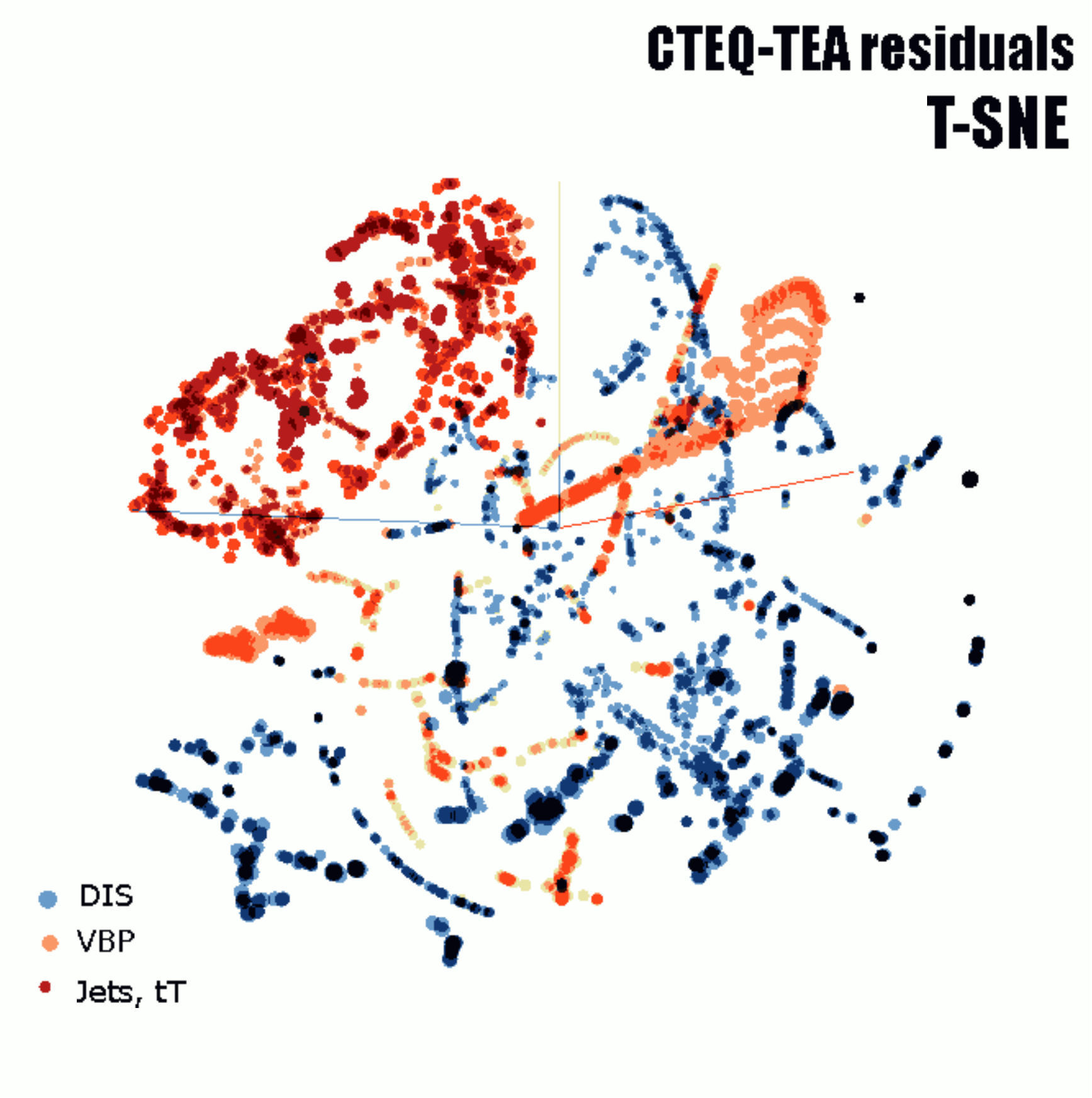}
\caption{Distributions of displaced residuals $\vec{\delta_{i}}$ from the
CTEQ-TEA analysis obtained by dimensionality reduction methods. Left:
a 3-dimensional projection of a 10-dimensional manifold constructed
by principal component analysis (PCA). Right: a distribution from
the 3-dimensional t-SNE clustering method. Blue, orange, and red colors
indicate data points from DIS, vector boson production, and jet/$t\bar{t}$
production processes, respectively. \label{fig:PCA-TSNE}}
\end{figure*}
}

\figone

\figtwo

\figthree

\null \vspace{-1.5cm}
\section{Introduction \label{sec:Introduction}}
\null \vspace{-0.8cm}

The determination of the nucleon's collinear parton distribution functions (PDFs)
is becoming an increasingly precise discipline with
the advent of high-luminosity experiments at both colliders and fixed-target
facilities, and several research groups are involved in the rich research
domain of modern PDF analysis.
%
PDFs provide a description of hadronic structure, and are an essential ingredient
in perturbative QCD computations.\footnote{%
See Ref.~\cite{Wang:2018heo} for additional details and a complete set of references.}
Since the start of the Large
Hadron Collider Run II (LHC Run II), the volume of experimental data
pertinent to the PDFs is growing with such speed that isolating measurements of
greatest impact presents a significant challenge for PDF fitters.
To help address this challenge, we present a new analysis method for
identifying high-impact experiments which constrain the PDFs and the
resulting Standard Model (SM) predictions that depend on them;
this method is complementary to other frameworks like Hessian
profiling techniques \cite{Camarda:2015zba} and Bayesian reweighting
\cite{Ball:2010gb,Ball:2011gg}.


\null \vspace{-0.8cm}
\section{Correlations}
\null \vspace{-0.8cm}

The notion of using correlations between
the PDF uncertainties of two physical observables was proposed in
Refs.~\cite{Pumplin:2001ct,Nadolsky:2001yg} as a means of quantifying
the degree to which these quantities were related based upon their
underlying PDFs. 
The PDF-mediated correlation $C_{f}$ can determine whether there \emph{may} exist a predictive relationship
between the PDF $f$ and goodness of fit to the $i^{th}$ data 
point; it  is defined as:\footnote{Here, 
the gradients $\vec{\nabla} f$ and r.m.s.\ sums $\Delta f$ are computed in 
the PDF parameter space.
}
\begin{equation}
  C_{f}\,\equiv\,\mbox{Corr}[f,r_{i}]
  =\frac{\vec{\nabla} f\cdot\vec{\nabla} r_{i}}{\Delta f\,\Delta r_{i}}
  \ .
  \label{eq:corr}
\end{equation}
We have suggestively inserted $f$ and $r_{i}$ as arguments of the correlation  function
where $f$ is a PDF and  $r_{i}$ is the residual constructed as $r_{i} = [T_i  - D^\mathit{sh}_{i}]\big/{s_i}$. 
We take $T_i$ as the theory prediction, $D^\mathit{sh}_{i}$ is the
datum shifted by the systematic uncertainties 
and $s_i$ is the uncorrelated uncertainty; 
see Ref.~\cite{Wang:2018heo} for a compete definition 
including the details of correlated uncertainties.

The Hessian correlation was deployed extensively in Ref.~\cite{Nadolsky:2008zw}
to explore implications of the CTEQ6.6 PDFs for envisioned LHC observables;
in this context it proved to be instrumental for identifying the specific PDF flavors
and $x$ ranges most tied to the PDF uncertainties for $W,$
$Z,$ $H$, and $t\bar{t}$ production cross sections as well as other
processes. 
%
%
%
At the same time, $C_{f}$ alone does not fully encode the potential
impact of measurements on improving PDF determinations
in terms of uncertainty reduction, particularly since the correlation of
Eq.~(\ref{eq:corr}) does not significantly depend on the {\it size} of experimental errors.

\null \vspace{-0.8cm}
\section{Sensitivity}
\null \vspace{-0.8cm}

As a remedy to these limitations,
we introduce a  generalization of the PDF-mediated correlations called
the \textit{sensitivity $S_{f}$}; this object
better identifies those experimental data points that tightly
constrain PDFs both by merit of their inherent precision and their
ability to discriminate among PDF error fluctuations. Such an approach
can aid in identifying regions of $\{x,\mu\}$ in which the PDFs are
particularly constrained by physical observables.

Thus, we define the \textit{sensitivity} $S_{f}$ to the PDF $f$ of the $i^{th}$
point in experiment $E$ to be: 
\begin{equation}
  S_{f}
  \equiv
  \frac{\vec{\nabla} f\cdot\vec{\nabla} r_{i}}{\Delta f\,\langle r_{0}\rangle_{E}}
  =
  \frac{\Delta r_{i}}{\langle r_{0}\rangle_{E}}\,C_{f}
  \ ,\label{eq:sens}
\end{equation}
where $\Delta r_{i}$  represents the variation of the
residuals across the set of Hessian error PDFs, 
 and we normalize it
to the r.m.s.\ residual for the experiment $E$  data-set,
$\langle r_{0}\rangle_{E}$, 
to reduce the impact
of random fluctuations in the data values.

This definition
has the benefit of encoding not only the correlated relationship of
$f$ with $r_{i}$, but also the comparative size of the experimental
uncertainty with respect to the PDF uncertainty. 
For
example, if new experimental data have reported uncertainties that
are much tighter than the present PDF errors, these data would then
register as high-sensitivity points.

In fact, in the numerical approach 
the user can quantify the sensitivity of data not only to individual
PDF flavors, but even to specific physical observables, including
the modifications due to correlated systematic uncertainties in every
experiment of the expanded CTEQ-TEA analysis. For example, for 14 TeV Higgs boson
production via gluon fusion ($gg\to H$) at the LHC, the short-distance
cross sections are known up to N$^{3}$LO with a scale uncertainty
of about 3\% \cite{Anastasiou:2016cez}. It has been suggested that
$t\bar{t}$ production and high-$p_{T}$ $Z$ boson production already
provide comparable constraints on the gluon PDF in the $x$ region
sensitive to LHC production, and that these are comparable to the
constraints from LHC and Tevatron data \cite{Czakon:2016olj,Boughezal:2017nla}.
Verifying the degree to which this hypothesis is true has been difficult
without actually including all these data in a fit.

As an alternative to doing a full global fit, we can critically assess
this supposition
using the Hessian correlations and sensitivities, $|C_{f}|$ and $|S_{f}|$,
associated with the Higgs production cross section $\sigma_{H^{0}}$,
in the context of the updated CTEQ-TEA set that includes the CT14HERA2~\cite{Hou:2016nqm}
points and newer LHC Run I data.

Fig.~\ref{fig:CorrSensH14} shows $|C_{f}|$ and $|S_{f}|$ distributions
that we obtain in $\{x,\mu\}$ space. The data points that have large
values of $|C_{f}|$ and $|S_{f}|$, and hence constrain the PDF dependence
of $\sigma_{H^{0}}$, are highlighted with color according to the
conventions described in Ref.~\cite{Wang:2018heo}. 
The sensitivity measure
generally identifies a different outlay of data providing constraints on $\sigma_{H^{0}}$
than the correlation, as can be seen by comparing highlighted
data points in the left and right panels.
Note, the thresholds are chosen to highlight the same effective number of points
in both plots. 
In the left panel, only a very small subgroup of the inclusive
jet production points, select HERA Neutral Current (NC) DIS measurements,
and several ATLAS $p^Z_T$ data, show the most significant correlations,
taken to have $|C_{f}|>0.42$ in this comparison. With our improved
definition for the sensitivity, however, the corresponding plot in
the right panel demonstrates that a different collection of points has large sensitivity
to $\sigma_{H^{0}}$, with $|S_{f}|>0.21$.
These data include
most of the analyzed jet production data, $t\bar{t}$ and high-$p_{T}$
$Z$ production, as well as various DIS experiments. 
From this comparison, one would conclude that efforts to constrain
PDF-based SM predictions for Higgs production relying only on a few
points of $t\overline{t}$ data would be significantly handicapped by
the neglect of high-energy jet production points.

\null \vspace{-0.8cm}
\section{Manifold learning and dimensionality reduction \label{subsec:Manifold-learning}}
\null \vspace{-0.8cm}



Lastly, we illustrate a possible analysis technique carried out with the help
of the TensorFlow Embedding Projector software for the visualization
of high-dimensional data \cite{Wang:2018heo,Cook:2018mvr}.
We operate on a table of 4021
vectors $\vec{\delta}_{i}$ defined from theory-data residuals according
to 
%
%
$ \vec{\delta}_{i} =
  {\delta}_{i,k} =
  [r_i(f_k) - r_i(f_0)]/ \langle r_{0}\rangle_{E} 
$
with $\langle r_{0}\rangle_{E}$ representing the {\it rms}-averaged
residual of the central fit evaluated over experiment $E$. This quantity can be computed
for the CTEQ-TEA data-set (corresponding to our total number of raw data points) and
is generated by our package \textsc{PDFSense} and uploaded to the Embedding Projector website.
As variations along many eigenvector directions result only in small
changes to the PDFs, the 56-dimensional $\vec{\delta}_{i}$ vectors
can in fact be projected onto an effective manifold spanned by fewer
dimensions. Specifically, the Embedding Projector approximates the
56-dimensional manifold by a 10-dimensional manifold using principal
component analysis (PCA). In practice, this 10-dimensional manifold
is constructed out of the 10 components of greatest variance in the
effective space, such that the most variable combinations of $\delta_{i,l}$
are retained, while the remaining 46 components needed to fully reconstruct
the original 56-dimensional $\vec{\delta}_{i}$ are discarded. However,
because the 10 PCA-selected components describe the bulk of the variance
of $\delta_{i,l}$, the loss of these 46 components results in only
a minimal relinquishment of information, and in fact provides a more
efficient basis to study $\delta_{i,l}$ variations.

In the 10-dimensional PCA representation, some directions result in
efficient separation of residuals of different types. For example,
the left panel of Fig.~\ref{fig:PCA-TSNE} shows a 3-dimensional
projection of the $\vec{\delta}_{i}$ that separates clusters of DIS,
vector boson production, and jet/$t\bar{t}$ production residuals.
In this example, the jet/$t\bar{t}$ cluster, shown in red, is roughly
orthogonal to the blue DIS cluster and intersects it. This separation
is remarkable, as it is based only on numerical properties of the
$\vec{\delta}_{i}$ vectors, and not on the meta-data about the types
of experiments that is entered after the PCA is completed.

As an alternative, the Embedding Projector can organize the $\vec{\delta}_{i}$
vectors into clusters according to their similarity using $t$-distributed
stochastic neighbor embedding (t-SNE). A
representative 3-dimensional distribution of the vectors obtained
by t-SNE is displayed in the right panel of Fig.~\ref{fig:PCA-TSNE}.
In this case, we find that such algorithms can again sort data into
clusters according to the experimental process, values of $x$ and
$\mu$, and even the experiment 
itself; for other examples, including animations, see Ref.~\cite{Cook:2018mvr}.


\null \vspace{-0.6cm}
\section{Conclusions \label{sec:Conclusions}}
\null \vspace{-0.6cm}

We have confronted the modern challenge of a rapidly growing set of
global QCD data with new statistical methodologies for quantifying and
exploring the impact of this information. These novel methodologies
are realized in a new analysis tool \textsc{PDFSense\cite{Wang:2018heo},}
which allows the rapid exploration of the
impact of both existing and potential data on PDF determinations.
Crucial to this analysis is introduction of the sensitivity $S_{f}$
which serves as a particularly powerful discriminator; both this and
the correlation $C_{f}$ allow us to visualize PDF constraints provided
by data across a wide range in $\{x,\mu\}$.
While we have demonstrated these techniques in the context of the
CT14 family of global fits, they are of sufficient generality that
one could readily repeat our analysis using alternative PDF sets.
These various tools collectively suggest
a number of possible avenues to advance PDF knowledge in the coming
years.

%
%
%
%
%

\bibliographystyle{unsrt}

\null

\end{document}